\documentclass[%
 reprint,
 amsmath,amssymb,
 aps,
prl,
floatfix,
]{revtex4-2}

\usepackage{graphicx}
\usepackage{dcolumn}
\usepackage{bm}
\usepackage{braket}
\usepackage{xcolor}

\begin{document}

\preprint{APS/123-QED}

\title{Self-injection-locked optical parametric oscillator based on microcombs}

\author{Fuchuan Lei$^{1,2}$}
\author{Yi Sun$^{1}$}
\author{Óskar B. Helgason$^{1}$}%
\author{Zhichao Ye$^{1}$}%
\author{Yan Gao$^{1}$}%
\author{Magnus Karlsson$^{1}$}%
\author{Peter A Andrekson$^{1}$}%
\author{Victor Torres-Company$^{1}$}%
 \email{torresv@chalmers.se}
\affiliation{$^1$Department of Microtechnology and Nanoscience, Chalmers University of Technology SE-41296 Gothenburg, Sweden
}%

\affiliation{$^2$ Key Laboratory of UV-Emitting Materials and Technology of Ministry of Ed, Northeast Normal University, Changchun, 130024, China}

\date{\today}

\begin{abstract}
Narrow-linewidth yet tunable laser oscillators are one of the most important tools for precision metrology, optical
atomic clocks, sensing, and quantum computing. Commonly used tunable coherent oscillators are based on stimulated
emission or stimulated Brillouin scattering; as a result, the operating wavelength band is limited by the gain media.
Based on nonlinear optical gain, optical parametric oscillators (OPOs) enable coherent signal generation within the
whole transparency window of the medium used. However, the demonstration of OPO-based Hertz-level linewidth and
tunable oscillators has remained elusive. Here, we present a tunable coherent oscillator based on a multimode coherent
OPO in a high-Q microresonator, i.e., a microcomb. Single-mode coherent oscillation is realized through self-injection
locking (SIL) of one selected comb line. We achieve coarse tuning up to 20 nm and an intrinsic linewidth down to sub-
Hertz level, which is three orders of magnitude lower than the pump. Furthermore, we demonstrate that this scheme results in the repetition rate stabilization of the microcomb. These results open exciting possibilities for generating tunable coherent radiation where stimulated emission materials are difficult to obtain, and the stabilization of microcomb
sources beyond the limits imposed by the thermorefractive noise in the cavity. 
\end{abstract}

\maketitle
Narrow-linewidth and tunable lasers are essential for applications that require low-phase-noise and wavelength versatility such as precision spectroscopy, quantum optics, optical atomic clocks, sensing, high performance communications and lidar, to name a few. The most widely used  approaches to implement narrow-linewidth lasers have relied on spectral purification of a free-running diode, solid-state or fiber laser, for example, active frequency locking the laser to a high-Q cavity with an electronic servo element \cite{webster2004subhertz,stoehr2006diode,notcutt2005simple,kessler2012sub} or optical feedback of a laser with an external optical element such as grating, mirror, cavity or waveguide \cite{liang2015ultralow,huang2016linewidth,shin2016widely,ji2022narrow,dang2023ultra}. Recent advances in photonic heterogeneous integration  have enabled coherent lasers with a linewidth at the 1 Hz level on a chip \cite{malik2021low}. The gain of most conventional narrow-linewidth lasers is mainly provided by stimulated emission, thus their  operating wavelength bands, i.e., maximum tuning ranges are fundamentally limited by the optical transitions allowed in the gain medium. In principle, narrow-linewidth and tunable coherent oscillators can be attained at extended wavelength regions by making use of nonlinear optical frequency conversion processes, such as harmonic generation, sum/difference frequency generation or four-wave mixing \cite{ling2023self,hill2022intra}. However, this remains a challenging endeavour because one or two narrow-linewidth pump lasers are required, and the phase matching condition needs to be stringently satisfied over a large frequency range. Stimulated Brillouin scattering is another well-known mechanism for implementing narrow-linewidth lasers \cite{debut2000linewidth,gundavarapu2019sub}, however, the emission wavelength is fundamentally constrained by the pump.

\begin{figure}[t]
\centering
\includegraphics[width=1\linewidth]{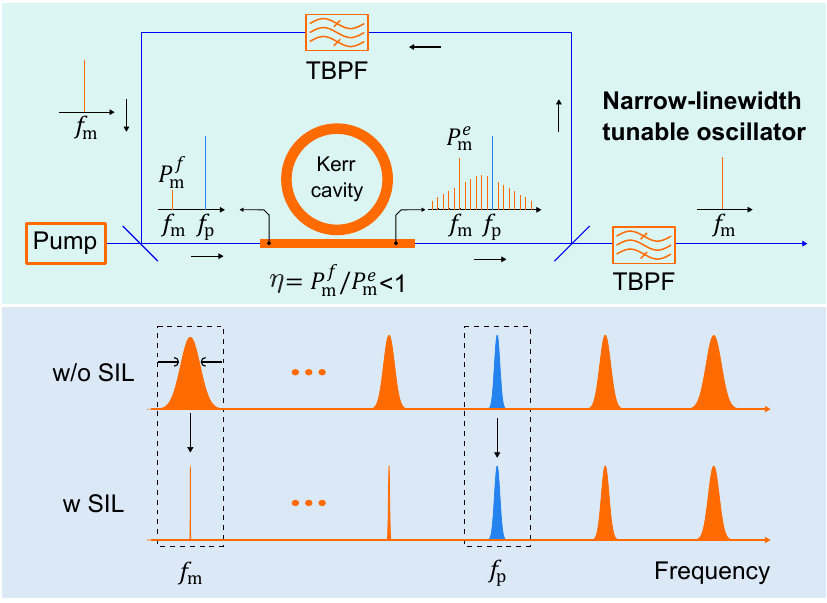}
\caption{Architecture of the self-injection-locked microcomb-based coherent oscillator. A microcomb is generated in a Kerr-nonlinearity microcavity with a continuous wave pump. The $m$-th comb line (counted from the pump, $f_{\rm m}$) is selected with a tunable bandpass filter (TBPF) and re-injected into the microcavity. With proper feedback power ratio ($\eta$), the linewidth of the $m$-th comb line (the coherent oscillator) is dramatically reduced, meanwhile, some other comb lines are narrowed accordingly, see Fig. \ref{fig:5}(a).}
\label{fig:1}
\end{figure}

Unlike ordinary lasers, the gain of optical parametric oscillators (OPOs) originates from an optically driven nonlinear polarization, thus the operating wavelength band is only limited by the absorption characteristics of the nonlinear medium used and the ability to engineer the phase matching among the waves involved in the process \cite{dunn1999parametric}. Nowadays, OPOs are widely exploited for tunable light generation ranging from UV to terahertz, particularly in the wavelength regions where ordinary lasers hardly reach \cite{breunig2011continuous}. The fundamental parametric process can be described as a new pair of photons, referred to as the signal photon(s)  ($\omega_{\rm s}$) and the idler photon(s) ($\omega_{\rm i}$), created from one (or two) pump ($\omega_{\rm p}$) photon(s) mediated by a ${\chi}^{(2)}$ (or ${\chi}^{(3)}$) nonlinear medium. According to energy conservation, the sum of the frequencies of signal and idler waves follows the pump frequency (or its double in the ${\chi}^{(3)}$ case), while their relative frequency could undergo a random diffusion process \cite{Graham1968}. Over the past decades, several approaches have been pursued in view of generating narrow-linewidth signal (or idler) via OPOs. The most commonly applied approaches include utilizing frequency-selective elements such as intracavity etalons or  gratings \cite{jacobsson2005narrowband,vainio2011tuning,peng2013high,das2015pump,saikawa2007high,zeil2014high,henriksson2007mid}. Self-injection seeding is an alternative technique for spectral narrowing in pulse pumped OPOs  \cite{rahm2004widely, xing2017widely}. These all-optical approaches permit signal spectral narrowing for several orders with compatibility of broadband wavelength tuning, however, the narrowing effect is usually insufficient to reach a signal wave with a linewidth smaller than the pump.
Another strategy is to stabilize the signal or idler to a frequency reference, such as a FP cavity \cite{ly201730,ricciardi2015sub,mhibik2010frequency,zhang2020cavity}, a frequency comb \cite{kovalchuk2005combination}, atomic
resonance \cite{zaske2010green}, or narrow-linewidth laser \cite{bae2013optical,zhang2020seeded}. These approaches could give superior spectral coherence but need a sophisticated servo system.

In this work, we present a narrow-linewidth and tunable coherent oscillator based on self-injection locking (SIL) of a multimode continuous-wave OPO, i.e. a Kerr microcomb \cite{del2007optical}. To be more specific, a microcomb with a single-soliton state is employed here \cite{herr2014temporal}. However, other coherent states such as Turing rolls or other soliton states \cite{xue2015mode,helgason2021dissipative}  could also be leveraged.   
The schematic architecture of the oscillator is shown in Fig. \ref{fig:1}. Like standard SIL configuration, the system is composed of a microcomb and an external optical feeback loop. To ensure a single-frequency narrow-linewidth oscillator is generated, a tunable bandpass filter (TBPF) is inserted into the feedback loop for selecting one comb line which can be arbitrary except for the pump. Optical gain can be involved into the feedback loop to compensate the loss, but in principle the feedback loop can be purely passive.
This configuration is thus fundamentally different from  laser cavity microcombs \cite{bao2019laser,selfsoliton2022,wang2016dual}, where the gain plays a crucial role for the comb generation. It is worth noting that, the terminology of SIL in microcombs has been
utilized in different contexts, including mode-locked state generation \cite{del2014self} and pump diode laser SIL
\cite{raja2019electrically,shen2020integrated,voloshin2021dynamics,taheri2022all}. OPO SIL has been discussed in the scenario of optical frequency division where the signal and the idler are subharmonics of the pump \cite{lee1999self,lee2003phase}. 
Here we explore and demonstrate the feasibility of dramatic spectral narrowing of a CW OPO with SIL, and realize the first sub-Hertz intrinsic linewidth oscillator based on parametric gain. 
We unveil a rich dynamics akin to what has been found in ordinary lasers with SIL \cite{petermann1991laser}, and show there exists a dynamic regime where the frequency noise of the comb line can be consistently reduced to be three orders lower than the pump, regardless of the feedback phase. Our work lays the foundation for understanding SIL dynamics in OPOs. In addition, this system allows us to select an arbitrary comb line (except for the pump) for SIL by simply setting the TBPF. Therefore, it enables simultaneous narrow-linewidth emission and continuous tuning in an extremely broad wavelength range. Furthermore, as the oscillator is one comb line of the soliton microcomb, its spectral purification can spontaneously facilitate microcomb stabilization beyond the limits imposed by the thermorefractive noise in the cavity \cite{drake2020thermal}.

\begin{figure*}[!t]
\centering
\includegraphics[width=0.95\linewidth]{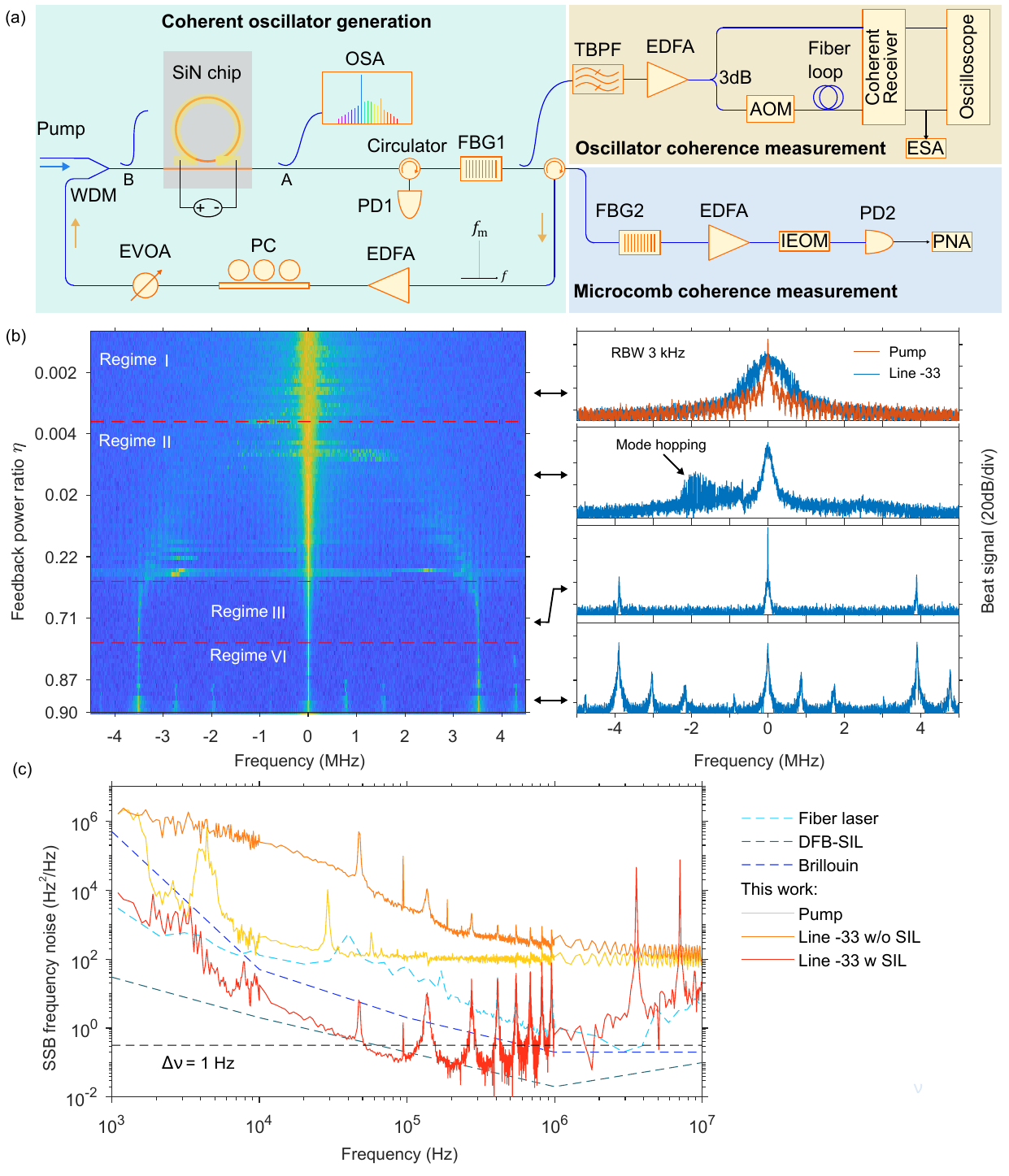}
\caption{Experimental study of the dynamics of the self-injection-locked microcomb-based oscillator. (a) Experimental setup. WDM, wavelength division multiplex. FBG, fiber Bragg grating. EDFA, erbium-doped amplifier. IEOM, intensity electro-optic modulator. ESA, electric spectrum analyzer. PNA, (ESA-based) phase noise analyzer. OSA, optical spectrum analyzer. AOM, acoustic optical modulator. EVOA, electronic variable optical attenuator. 
FBG1 is applied for pump (1536.4 nm) rejection while FBG2 is for selecting one comb line (1562.8 nm) for SIL. The 3 dB bandwidth of the FBG filters is $\sim$100 GHz and their center-wavelength tuning range is $\sim$10 nm. (b) The evolution of the laser (comb line -33) delayed heterodyne beating spectra with feedback power ratio. The beating spectrum of the pump is also plotted as a reference. (c) Single sideband (SSB) frequency noise of the oscillator in this work and three other types of lasers. The data for DFB-SIL and Brillouin laser are taken from refs \cite{li2021reaching} and \cite{gundavarapu2019sub}. As a base line, the frequency noise corresponding to 1 Hz Lorentzian linewidth is given  \cite{di2010simple}.}
\label{fig:2}
\end{figure*}

\section{Experimental study of SIL dynamics}

In this section, we present the experimental study of the SIL dynamics.  The setup is shown in Fig. \ref{fig:2}(a). An integrated silicon nitride ($\rm {SiN}$) microresonator is used for microcomb generation. The height and the width of the SiN waveguide are 740 nm and 1800 nm, respectively, which result in a group velocity dispersion coefficient of $\beta_2=-70$ $ {\rm ps}^2/{\rm km}$ for the $\rm {TE}_{00}$ mode. The radius of the microresonator is 227 $\mu $m, so the corresponding FSR is $\sim$100 GHz. The gap between ring and bus is 450 nm. Both the average intrinsic and external quality factors for $\rm {TE}_{00}$ mode family are $\sim9\times 10^6$.  Two lensed fibers are used for light coupling into and out of the on-chip SiN bus waveguide. The coupling loss per facet is $\sim$2 dB. A narrow-linewidth  external cavity diode laser ($\lambda=1536.9$ nm) after being amplified by an erbium-doped fiber amplifier (EDFA) is employed as the pump. A single-soliton microcomb is generated at the $\rm {TE}_{00}$ mode family via fast thermo-optic tuning. In this initial experiment, we select the comb line -33 ($\lambda=1562.8$ nm) by means of a fiber brag grating (FBG) filter for SIL, but in principle any other mode could be used. 

To investigate the SIL dynamics, an EDFA (small-signal gain $\sim$35 dB) and an electronic variable optical attenuator (EVOA) are included into the feedback loop for feedback strength control. To make the feedback to be efficient, a fiber polarization controller  is inserted into the feedback loop. The optical loss of the feedback loop is around 10 dB when the EDFA is off. The total length of the feedback fiber loop is $\sim$50 m, including $\sim$27 m fiber in the EDFA. The feedback strength can be quantified as the on-chip power ratio of the feedback field re-entering the microresonator ($P^{f}_{\rm m}$) and  emitting one ($P^{e}_{\rm m}$), i.e., $\eta= {P^{ f}_{\rm m}}/{P^{ e}_{\rm m}}$, see Fig. \ref{fig:1}. In the experiment, $\eta$ can be measured from the comb line power ratio at nodes A and B (Fig. \ref{fig:2}) when the feedback loop is closed. It is important to emphasize that $\eta <1$ is satisfied all the time in this work, see Fig. \ref{fig:2}(b), that is to say, the EDFA here is not employed as the 'gain source' for the compound cavity laser but instead it is served as the loss compensator of the feedback loop. The  equivalent Q-factor of the feedback loop is $\sim-3\times 10^8/{\rm ln}\eta$. Akin to opto-electronic oscillator, the fiber feedback loop could potentially be replaced by a high-Q microresonator \cite{OEO}.

\begin{figure*}[t]
\centering
\includegraphics[width=0.9\linewidth]{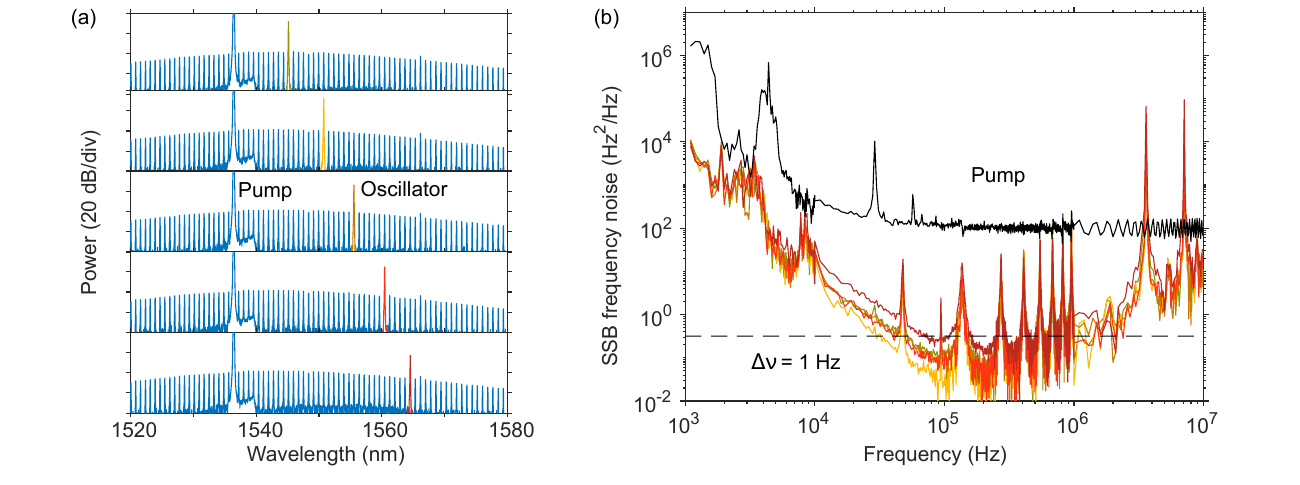}
\caption{Wavelength tunability demonstration. The narrow-linewidth oscillators were realized at different wavelengths via selecting different comb lines for SIL. (a) Optical spectra. (b) The corresponding frequency noise PSDs.}
\label{fig:3}
\end{figure*}

To analyze the spectral characteristics of the single-mode coherent oscillator, the selected comb line is amplified and filtered for out-of-loop characterization. Subsequently, the oscillator is divided into two paths by a 3 dB coupler. One beam is frequency shifted by an AOM (80 MHz) and then delayed through a $\sim$1.5 km single-mode fiber. The two beams are then sent into a coherent receiver. As a result, a pair of delayed self-heterodyne beating signals with center frequency 80 MHz are obtained, which can be recorded for extracting the frequency noise power spectral density (PSD) of the oscillator. To get a qualitative evaluation of the spectral purity, this beating signal is also directly monitored by an ESA.

Figure \ref{fig:2}(b) shows the beating signal spectrum evolves with the feedback strength. From top to bottom, $\eta$ increases monotonically (in a nonlinear manner for better display), by tuning the EVOA. Considering the phase of the fiber feedback loop is not a constant because of ambient temperature fluctuation \cite{ma1994delivering}, here we do not investigate the dynamics with a specific feedback phase since it is hard to measure and control in practice.

Similar to semiconductor lasers with optical feedback \cite{petermann1991laser}, this system contains a wealth of nonlinear dynamics. We found the dynamics of this system can also be classified as four distinct feedback regimes \cite{tkach1986regimes,schunk1988numerical} according to the comb line's noise performance, see Fig. \ref{fig:2}(b).
Regime $\rm \uppercase\expandafter{\romannumeral1}$ corresponds to very weak feedback strength. The spectral width can be slightly broadened or narrowed from time to time, inferred as the variation of the feedback phase. With increasing feedback strength, spectral line narrowing is observed most of time. However, mode-hopping starts to occur frequently with a given feedback strength, reflected as a burst of spikes in the beating spectrum, see Fig. \ref{fig:2}(b). This regime is termed as regime $\rm \uppercase\expandafter{\romannumeral2}$. 

When the feedback reaches to a certain level (regime $\rm \uppercase\expandafter{\romannumeral3}$), the mode hopping ceases altogether and the line is persistently narrowed. This means the spectral narrowing is no longer sensitive to the feedback phase. Therefore, this regime allows to obtain an extremely narrow-linewidth oscillator under relaxed conditions. In the following, the SIL state refers to this regime unless otherwise specified.

If the feedback strength is increased further (regime $\rm \uppercase\expandafter{\romannumeral4}$), a so-called coherence-collapsed state is obtained \cite{tkach1986regimes,schunk1988numerical}. Not only the original beating note spectrum gets broadened, but new frequency components emerge. If the feedback ratio goes close to unit, the feedback power is too strong and the soliton would vanish. A further detailed discussion of the dynamics would be lengthy and beyond the scope of this work. In this work, we mainly focus on the discovered regime $\rm \uppercase\expandafter{\romannumeral3}$.

\begin{figure*}[!t]
\centering
\includegraphics[width=1\linewidth]{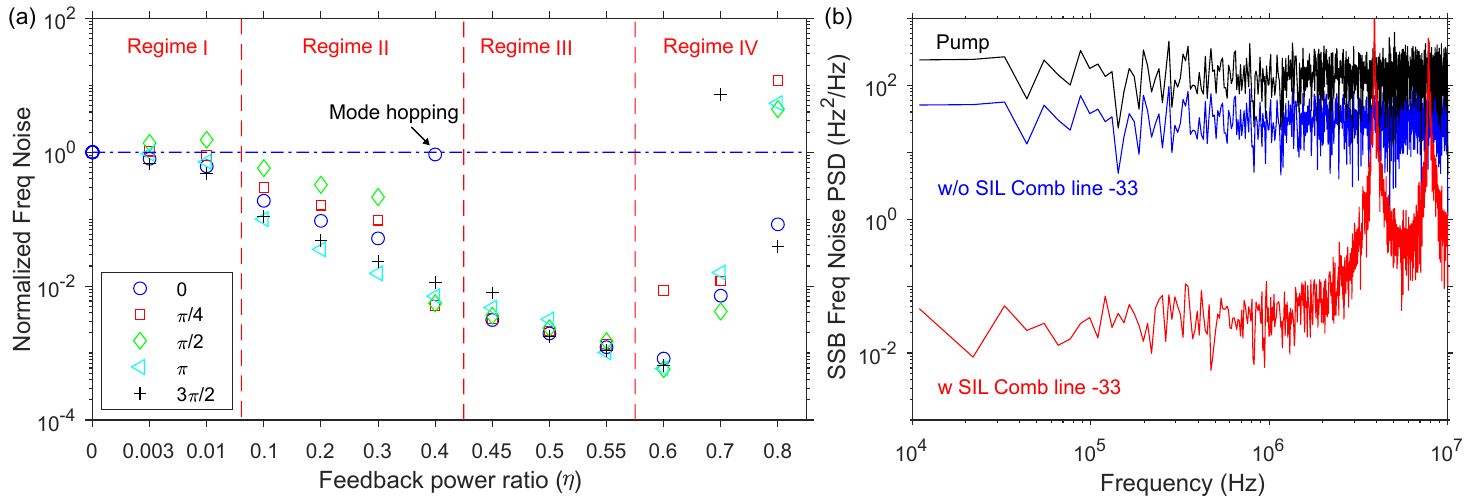}
\caption{Numerical study of SIL of soliton microcomb. (a) Frequency noise reduction (or amplification) as a function of feedback strength and phase. (b) Comparison of frequency noise PSD.}
\label{fig:4}
\end{figure*}

Figure \ref{fig:2}(c) shows the measured frequency noise 
of the pump and comb line -33 for the cases with and without SIL, which is measured by the correlated self-heterodyne method \cite{yuan2022correlated}. For the case without SIL, comb line -33 features a much higher frequency noise than the pump mainly due to the thermorefractive noise of the microcavity \cite{huang2019thermorefractive,drake2020thermal,lei2022thermal}. However, its frequency  noise got reduced by three to four orders of magnitude after SIL, resulting in a sub-Hertz intrinsic linewidth, which is two orders lower than the pump. The results indicate both the pump noise and thermorefractive noise can be greatly suppressed through SIL. Although it is hard to provide an analytical model for understanding the SIL dynamics in this multimode nonlinear system, particularly in regime $\rm \uppercase\expandafter{\romannumeral3}$, the noise reduction through SIL can be intuitively explained as large cavity induced noise dilution \cite{kazarinov1987relation}.   

The spectral coherence of the oscillator is comparable to other types of narrow-linewidth lasers, i.e. fiber laser (NKT Photonics, Koheras), self-injection locked DFB laser \cite{li2021reaching} and Brillouin laser \cite{gundavarapu2019sub}, see Fig. \ref{fig:2}(c). 
As a common feature of feedback, the beating spectrum as well as the optical frequency noise PSD exhibits peaks at frequencies close to the multiples of the FSR (~3.6 MHz) of the feedback loop. However, these spikes can be greatly suppressed by the microresonator if their frequencies are tens of MHz, i.e.,  a meter-level long feedback loop is utilized. 

The above demonstrated self-injection-locked microcomb can be easily explored as a wavelength-tunable coherent oscillator. The emission wavelength could be selected by using different comb line for SIL. As shown in Fig. \ref{fig:3}, we demonstrate the Hertz-level oscillator is not exclusively attainable for a specific comb line. Instead it can be applicable to any comb lines with wavelength ranging from 1545 nm to 1565 nm, limited by the working range of our filters. Although not shown here, it should be possible to obtain continuous wavelength tuning via simultaneous tuning the pump frequency and the cavity resonance over one FSR of the microresonator.  Nevertheless, the long-term stability of the coherent oscillator cannot be guaranteed, because the whole system is operated in atmospheric environment without any servo control loop. This issue could be solved in the future with a better packaged device or better yet with a fully integrated system. Further improvements could be attained by locking the oscillator to a stable passive optical reference, such as on-chip stable cavities \cite{Guochip1Hz}.

\section{Numerical study of SIL dynamics}
For a better understanding of the SIL dynamics of the system, we performed a numerical study. Different from semiconductor lasers where the noise is dominated by spontaneous emission, the noise sources of the current system include pump phase and intensity noise, thermorefractive noise of the microresonator, shot noise and technical noise \cite{lei2022optical}. To grasp the essential dynamics and facilitate understanding, we only consider the pump phase noise here. The other noise sources can be included through a similar approach. The detailed simulation method is described in the Supplementary Materials. 

Firstly, we numerically investigate the SIL dynamics at different feedback strengths. In this case, the pump phase noise is simplified as a single tone modulation of the pump phase at 1.2 MHz. Figure \ref{fig:4}(a) shows the simulated frequency noise induced by pump phase noise, which is normalized to the case without feedback, i.e., $\eta=0$. A variety of feedback strengths and feedback phases are considered. Consistent with the experimental observations, four dynamics regimes can be clearly distinguished according to the noise reduction performance and sensitivity to the feedback strength. In regime $\rm \uppercase\expandafter{\romannumeral1}$, the noise could be reduced or enlarged according to feedback phase.  In regime $\rm \uppercase\expandafter{\romannumeral2}$, the noise is usually suppressed but the  suppression ratio is phase dependent, and mode hopping can also be observed. With increasing feedback strength, the aforementioned regime $\rm \uppercase\expandafter{\romannumeral3}$ is replicated, where the noise reduction is nearly independent of the feedback phase.
In $\rm \uppercase\expandafter{\romannumeral4}$, the system starts to enter chaotic states or remains at stationary state according to the feedback phase, and with increasing feedback strength, it becomes completely chaotic regardless of the feedback phase.

Secondly, we performed a frequency noise PSD simulation comparing the system operated at the regime  $\rm \uppercase\expandafter{\romannumeral3}$ and without feedback case. The frequency noise of the pump is modeled as a flat PSD, i.e., white noise. 
It is noted that comb line -33 features a slightly lower frequency noise than the pump without SIL, which can be attributed to the Raman effect \cite{lei2022optical}. After SIL, the frequency noise is reduced by three orders at low-offset region, see Fig \ref{fig:4}(b). The characteristic spikes caused by feedback can also be seen, indicating that the numerical simulation captures the key features of the SIL dynamics. In the simulation, we only considered a constant feedback power ratio and did not include the gain dynamics of the EDFA, indicating the essential role of the EDFA used in the feedback loop is a loss compensator.

\begin{figure}[!b]
\centering
\includegraphics[width=1\linewidth]{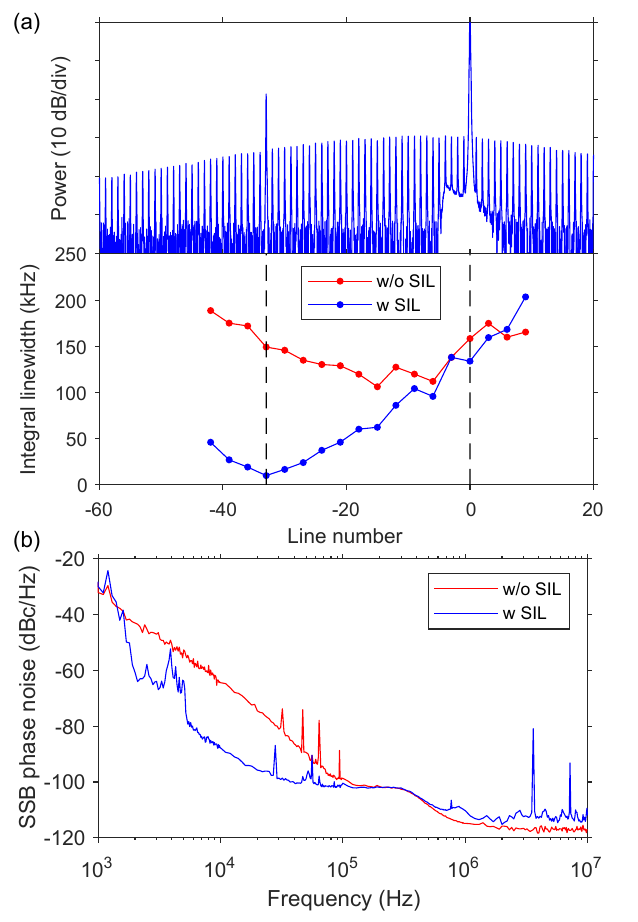}
\caption{Coherent oscillator facilitated soliton microcomb stabilization. (a) The optical spectrum of the single-soliton microcomb with SIL and the measured integral linewidth  of comb lines according to the frequency noise PSD. (b) Phase noise of the down converted repetition rate the soliton microcomb.}
\label{fig:5}
\end{figure}

\begin{figure}[!t]
\centering
\includegraphics[width=1\linewidth]{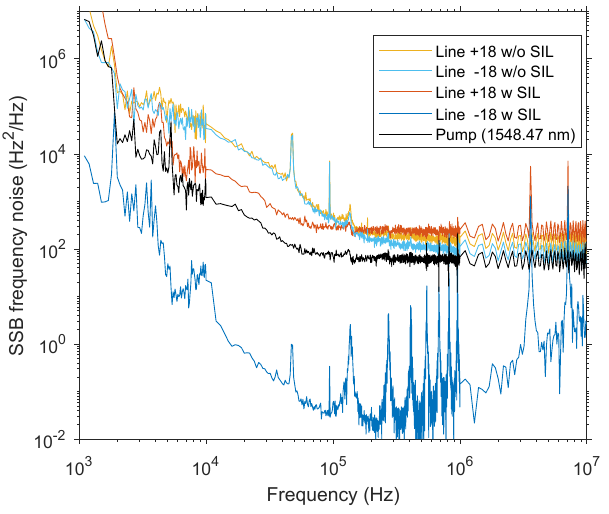}
\caption{Frequency noise of a pair comb lines   symmetrically located around the pump, with one side being selected for SIL.}
\label{fig:6}
\end{figure}

\section{Coherent Oscillator facilitated Soliton microcomb stabilization}
The above narrow-linewidth oscillator is one comb line of the soliton microcomb, therefore its phase should be highly correlated with the other comb lines according to the elastic tape model \cite{lei2022optical}. The measured integral linewidth ($\beta$-line separation algorithm \cite{di2010simple}) of the comb lines for the cases with and without SIL are compared and presented in Fig. \ref{fig:5}(a). Not surprisingly, the pump linewidth is not changed because it is predetermined by the external pump source, however, the other comb lines get an effective linewidth reduction benefited from the SIL of one comb line, particularly for its neighbor lines. This phenomenon can be understood by noting that, in a frequency comb, there are only two degrees of freedom.  Recent works are actively investigating the stabilization of microcombs using two low noise pumps by optical frequency division \cite{ mmThz, taheri2022all, SSIL, Kudelin,sun}.  Here, the comb line with SIL is derived from the microcomb, and together with the pump  determine the two degrees of freedom of the comb. 

Besides the spectral measurement of comb lines, we also measure the stability of the repetition rate. For this purpose, the repetition rate is electro-optic downconvered into low frequency range ($\sim$ 470 MHz) with a 25 GHz intensity modulation, and then its phase noise is measured with a phase-noise analyzer. As shown in  Fig. \ref{fig:5}(b), its single-sideband phase noise is reduced at low-offset frequencies while increased at higher frequency offset ($>$100 kHz) after SIL. This phenomenon can be understood as the repetition rate phase noise is mainly determined by the  relative frequency stability 
between the pump and the narrow-linewidth oscillator.  At lower offset frequencies, the relative frequency noise is mainly limited by the linewidth of the comb line due to the thermorefractive noise, therefore, comb linewidth narrowing can lead to lower repetition rate phase noise. While, at higher offset frequency region, the linewidth of the pump becomes to be a limiting factor as comb line -33 features a much lower frequency noise (or higher at the spike frequencies). 

To make it clear, we change the pump wavelength to 1548.47 nm and select comb line -18 for SIL, making its mirror comb line, i.e., line 18 placed within our measurable range (C-band).  As shown in Fig. \ref{fig:6}, the frequency noise of comb line 18 is reduced at low-offset frequencies while increased at higher frequency offset after SIL. Hence, in order to exploit the narrow-linewidth oscillator for extremely pure microwave generation, a narrower linewidth pump laser is required. 

\section{Discussion}
 In summary, we have demonstrated a narrow-linewidth yet tunable oscillator through SIL of a single comb line from a soliton microcomb.  A dynamic regime suitable for  parametric oscillator SIL is discovered both experimentally and numerically. Considering the gain of microcombs is based on parametric amplification, and octave-spanning spectra can be directly generated in a monolithic nonlinear microresonator \cite{del2011octave,chen2020chaos}, our approach hence has potential to generate extremely coherent electromagnetic radiation at wavelength regions where attaining optical amplification by stimulated emission of radiation is challenging, while using pump lasers in the near infrared.
Due to the existence of fiber-chip coupling loss and filter insertion loss, an optical amplifier is still utilized in the feedback loop, however, these losses could be overcome by developing a fully integrated low loss photonic delay line \cite{lee2012ultra,wang2017continuously,lin2022high} or a high-Q  microring filter. In terms of the output power, owing to the low conversion efficiency of the single-soliton microcomb, the power of this laser currently is limited to sub-milliwatt (0.2 mW) level, which can be boosted further with other types of OPOs featuring much higher output power \cite{lei2023hyperparametric,helgason2022power} or parametric amplifiers \cite{ye2021overcoming,riemensberger2022photonic,geng2019microcavity}. Moreover, our scheme should  be also applicable to ${\chi}^{(2)}$ media based OPOs, which usually feature even higher conversion efficiency and output power compared to the ${\chi}^{(3)}$ case.

In addition to the realization of a single-tone pure oscillator, we also demonstrate this scheme has the potential for thermorefractive noise suppression and microcomb stabilization, critical for ultralow noise microwave generation. 

\section{Appendix}
The microcomb noise dynamics simulation is carried out with the Ikeda map method.
A full roundtrip evolution is composed of two steps: (1) the coupling between bus waveguide and microresonator, (2) nonlinear propagation in the microresonator over its circumference.  To consider the feedback effect given by the feedback loop and tunable bandpass filter, a feedback field is added into the input field.

The coupling between bus waveguide and resonator can be described as 
\begin{eqnarray}
A_{\rm n+1}(0,\tau)=\sqrt{\theta}(A_{\rm n}^{\rm in}+A^{\rm f}_{\rm n})+\sqrt{1-\theta}e^{i\phi_0}A_{\rm n}(L,\tau),\\
A^{\rm out}_{\rm n}(0,\tau)=\sqrt{1-\theta}(A_{\rm n}^{\rm in}+A^{\rm f}_{\rm n})-\sqrt{\theta}e^{i\phi_0}A_{\rm n}(L,\tau),
\end{eqnarray}
where $A_{\rm n}$ stands for the amplitude (normalized to power) of intracavity field of $n$-th round trip, $\phi_0 ={2\pi(\nu_{\rm p}- \nu_{\rm c})}/{\rm FSR}$, and $\theta = {2\pi \nu_{\rm p}}/({{\rm FSR}\times Q_{\rm ex}}) $ with $Q_{\rm ex}$ as the extrinsic quality factor. $\nu_{\rm p}$ and $\nu_{\rm c}$ stand for the pump frequency and cavity resonant frequency of the pump mode.  

The feedback field $A^{\rm f}_{\rm n}$ is given by the modified delayed output:
\begin{equation}
  { A^{\rm f}_{\rm n}}= 
\left\{ 
    \begin{array}{lc}
      0 & n \leq l \\
        \mathcal{F}^{-1}\{\sqrt{\eta} \mathcal{F}[A^{\rm out}_{n-l})](\mu)e^{-i\phi_{\rm f}}\}&n> l\\
   \end{array},
\right.
\end{equation}
where $\mathcal{F}$ stands for Fourier transform, $\mu$ denotes the mode index for the comb line used for SIL in the frequency domain, $l/{\rm FSR}$ and $\phi_{\rm f}$ represent the roundtrip time and the phase delay of the feedback loop. In the simulation, the feedback loop was set to be 50 m.

At each  propagation step, the generalized nonlinear Schrödinger equation with inclusion of Raman scattering is solved:
\begin{eqnarray}
\begin{split}
    \frac{\partial A}{\partial z}+ \frac{\alpha}{2} A+i\frac{\beta_{\rm 2}}{2}\frac{\partial^2 A}{\partial t^2}  = i\gamma A(z,t) \times  
    \int_0^\infty R(t')|A(z,t-t')|^2dt'.  \nonumber
\end{split}
   \end{eqnarray}
Here $R(t)=(1-f_{\rm R})\delta(t)+f_{\rm R} h_{\rm R}(t)$, and $h_{\rm R}(t)=(\tau_1^{-2}+\tau_2^{-2})\tau_1 {\rm exp}({-t}/{\tau_2}){\rm sin}({t}/{\tau_1})$. In the simulation, the parameters $\tau_1=15 $ fs, $\tau_2=120 $  fs, $f_{\rm R}=0.03$ are  used as they match the measured optical spectrum. The remaining  parameters  are directly measured or calculated and have the following values: $
P_{\rm in}=0.16$ W,
$\gamma=0.98 {\rm W^{-1}m^{-1}}$, $Q_{\rm ex}=9\times10^6$, $Q_{\rm in}=9\times10^6$, $\nu_{\rm c}-\nu_{\rm p}= 500$ MHz.

In the study of SIL dynamics (Fig. 4(a)), the pump phase noise is included into $A_{\rm n}^{\rm in}$ by assigning it a time-dependent  phase $\phi_{\rm n}$ as
\begin{equation}
    \phi_{\rm n}=\epsilon {\rm sin}(n \Omega_{\rm p}/ {\rm FSR}),
 \end{equation}
 where $\epsilon$ stands for the phase modulation depth, and $\Omega_{\rm p}=1.2$ MHz.
To simulate the frequency noise PSD (Fig. 4(b)), the pump phase noise is set to be [1]
\begin{equation}
    \phi_{\rm n+1}=\phi_{\rm n}+\sqrt{2\pi\Delta\nu_{\rm p}/{\rm FSR}}\times\xi,
 \end{equation}
where $\Delta\nu_{\rm p}$ and $\xi$ stands for the intrinsic linewidth of the pump and a normally distributed random number, respectively.

As indicated by Eq. (3), the simulation program is first started for the soliton microcomb without feedback. The values for the feedback field are stored in a memory and it gets reused after the roundtrip time of the feedback cavity in order to account for the optical feedback. The computer program has run for another  5 million roundtrips to achieve a 'stationary' description of the laser with optical feedback. The phases of the select mode for SIL were recorded every 512 roundtrips. Based on the recorded phases, the frequency noise PSD was computed.

 The SiN devices demonstrated in this work were fabricated at Myfab Chalmers.

\bibliographystyle{apsrev4-2}
\bibliography{ref}

\end{document}